\documentclass[twocolumn]{aastex63}

\newcommand{\fluxu}{erg s$^{-1}$ cm$^{-2}$ \AA$^{-1}$}

\newcommand{\HeHr}{\ion{He}{2}($\lambda4686$)/H$\beta$}

\newcommand{\NW}{HT15}
\newcommand{\SE}{HT5}
\newcommand{\MB}{CHR}

\shorttitle{Izw18 Letter}
\shortauthors{Rickards Vaught et al.}
\graphicspath{{./}{figures/}}
\begin{document}
\title{Keck Cosmic Web Imager Observations of He II Emission in I Zw 18}
\correspondingauthor{Ryan Rickards Vaught}
\email{rjrickar@ucsd.edu}

\author{Ryan J. Rickards Vaught}
\affiliation{Center for Astrophysics and Space Sciences, Department of Physics, University of California, San Diego, 9500 Gilman Dr., La Jolla, CA 92093, USA}

\author{Karin M. Sandstrom}
\affiliation{Center for Astrophysics and Space Sciences, Department of Physics, University of California, San Diego, 9500 Gilman Dr., La Jolla, CA 92093, USA}

\author{Leslie K. Hunt}
\affiliation{INAF-Osservatorio Astrofisico di Arcetri, Largo E. Fermi 5, I-50125 Firenze, Italy}

\begin{abstract}
With a metallicity of 12+Log(O/H) $\approx 7.1-7.2$, I Zw 18 is a canonical low metallicity blue compact dwarf (BCD) galaxy. A growing number of BCDs, including I Zw 18, have been found to host strong,  narrow-lined, nebular \ion{He}{2} ($\lambda$4686) emission with enhanced intensities compared to H$\beta$ (e.g. \HeHr\ $>1\%$). We present new observations of I Zw 18 using the Keck Cosmic Web Imager. These observations reveal two nebular \ion{He}{2} emission regions (or \ion{He}{3} regions) northwest and southeast of the \ion{He}{3} region in the galaxy's main body investigated in previous studies. All regions exhibit \HeHr\ greater than $2\%$. The two newly resolved \ion{He}{3} regions lie along an axis which intercepts the position of I Zw 18's Ultra-luminous X-ray (ULX) source. We explore whether the ULX could power the two \ion{He}{3} regions via shock activity and/or beamed X-ray emission. We find no evidence of shocks from the gas kinematics. 
If the ULX powers the two regions, the X-ray emission would need to be beamed. Another potential explanation is that a class of early-type Nitrogen-rich Wolf-Rayet stars with low winds could power the two \ion{He}{3} regions, in which case the alignment with the ULX would be coincidental.
\end{abstract}


\section{Introduction}\label{sec:intro}
Narrow-lined, nebular \ion{He}{2}($\lambda4686$) emission (\ion{He}{2} emission) is observed in an increasing number of Blue Compact Dwarf (BCD) galaxies. Emission from \ion{He}{2} arises from the recombination of doubly-ionized Helium, which requires energetic photons $>$ 54 eV. Although there have been several studies into the origin of \ion{He}{2} emission  \citep{Garnett1991ApJ...373..458G,Izotov1998ApJ...497..227I,Cervino2002A&A...392...19C,Thuan2005ApJS..161..240T,Kehrig2011A&A...526A.128K,Shirazi2012MNRAS.421.1043S,Kehrig2015ApJ...801L..28K,Kehrig2018MNRAS.480.1081K,Schaerer2019A&A...622L..10S,Senchyna2020MNRAS.494..941S}  the source(s) of the required ionizing flux remains uncertain.

Of the BCDs that exhibit \ion{He}{2} emission, I Zw 18 is of unique interest as it is relatively nearby \citep[$18.2\pm 1.5$ Mpc;][]{Aloisi2007}, corresponding to a distance modulus of 31.3 Mag, and has one of the lowest metallicities, 12+Log(O/H) $\approx 7.1-7.2$ \citep[Rickards Vaught et al. in prep,][and references therein]{Searle1972ApJ...173...25S,Izotov1999ApJ...511..639I,Kehrig2016MNRAS.459.2992K}. \ion{He}{2} emission in I Zw 18 has a history of being observed via single-slit spectroscopy \citep{Bergeron1977ApJ...211...62B,Garnett1991ApJ...373..458G,1997ApJ...487L..37I,Izotov1998ApJ...497..227I,1998ApJ...508..248V} which has the disadvantage of sparse spatial sampling. Recently, integral field spectroscopy (IFS) of I Zw 18 has spatially resolved the extent of \ion{He}{2} emission near the NW stellar cluster \citep{Kehrig2015ApJ...801L..28K}.

I Zw 18 also hosts an X-ray Binary (XRB), near this \ion{He}{3} region. The first reported X-ray luminosity, $L_X$, of this XRB, via \textit{Chandra} imaging \citep{Bomans2002ASPC..262..141B}, was $\sim 10^{39}$ erg s$^{-1}$ in the $0.5-10$ keV band \citep{Thuan2005ApJS..161..240T}. \cite{Kehrig2015ApJ...801L..28K} modeled the XRB contribution to the \ion{He}{2} luminosity, $L_{\lambda 4686}$,
and found that the predicted $L_{\lambda 4686}$ is $\sim 100\times$ weaker then their observed value. However, deeper \textit{XMM-Newton} imaging and analysis of the XRB by \cite{Kaaret2013ApJ...770...20K} report a $0.3-10$ keV band luminosity, $L_X=1.4 \times$ 10$^{40}$ ergs s$^{-1}$, with a harder spectrum than observed with \textit{Chandra}. With $L_X$ $> 10^{39}$ ergs s$^{-1}$ this XRB is considered an Ultra-luminous X-ray (ULX) source \citep{Pakull2006IAUS..230..293P,Kaaret2013ApJ...770...20K}. 
Although the best fit to the \textit{XMM-Newton} spectrum, assuming sub-Eddington accretion, suggests a black hole with mass $> 154 \text{ M}_{\odot}$ \citep{Kaaret2013ApJ...770...20K},
recent work suggests that a significant fraction of ULXs are instead stellar-mass binary systems undergoing super-Eddington accretion \citep[][and references therein]{King2020MNRAS.494.3611K}. 
Recently, infrared observations and photoionization modeling by \cite{Lebouteiller2017A&A...602A..45L} suggest that I Zw 18's neutral gas heating can be explained by this single XRB, if the $10^{4}$ yr time averaged $L_X$ is 4 $\times$ 10$^{40}$ ergs s$^{-1}$.

In our deeper, higher angular and velocity resolution Keck Cosmic Web Imager (KCWI) observations, we detect two additional \ion{He}{3} regions in I Zw 18. These regions are NW and SE of the emission reported by \citet{Kehrig2015ApJ...801L..28K}, and lie along an axis that intercepts the position of I Zw 18's ULX. The alignment may be coincidental or may suggest that the ULX powers these two regions. 
We describe our observations and data reduction in Section \ref{sec:obs}. Our emission line fitting is detailed in Section \ref{sec:velfit}. Section  \ref{sec:astrometry} outlines how we determine the position of the ULX source. We present our results in Section \ref{sec:results}. We discuss possible ionizing sources of the two newly resolved \ion{He}{3} regions in Section \ref{sec:discussion}. We conclude this letter in Section \ref{sec:conclusions}.

\section{Observations and Data Analysis}\label{sec:obs}
\subsection{Archival Data}
Several archival datasets are used in the course of analyzing our KCWI observations. We downloaded archival Hubble Space Telescope\footnote{Based on observations made with the NASA/ESA Hubble Space Telescope, and obtained from the Hubble Legacy Archive, which is a collaboration between the Space Telescope Science Institute (STScI/NASA), the Space Telescope European Coordinating Facility (ST-ECF/ESA) and the Canadian Astronomy Data Centre (CADC/NRC/CSA).} (\textit{HST}) images in the F439W and F814W filters (Program ID: 5434, 10586) as well as \textit{r}-band Sloan Digital Sky Survey \citep[SDSS,][]{SDSS2003AJ....126.2081A} images of I Zw 18 (Fields: 157, 158, 238, and 239). To compare with the ULX in I Zw 18, we also downloaded \textit{Chandra} X-ray imaging of the galaxy \citep{Bomans2002ASPC..262..141B}. 

\subsection{KCWI Observations and Data Reduction}
The IFS data were taken in clear conditions on December 25th, 2017 using KCWI installed on the 10-meter Keck II Telescope. We used the small slicer and BL grating centered at 4550 \AA\ with a usable spectral range of 3700-5500 \AA. The spectral resolution, R$\sim$3600 corresponds to a full-width-half-max (FWHM) $\sim 1.26$ \AA\, at 4550\AA. The slice width is 0.35\arcsec. Each pointing covers a field of view (FoV) 8.5\arcsec\ perpendicular and 20.4\arcsec\ parallel to the slicer. Using images of the standard star Feige 34, taken in the same conditions, we measured the FWHM of the point spread function to be $\sim 0.7 \arcsec$.

The main body of I Zw 18 comprises two stellar clusters \citep[IZW18-NW and IZW18-SE,][]{Skillman1993ApJ...411..655S}, shown in Figure \ref{fig:regions}, and is not covered by a single instrumental FoV. To cover the galaxy, we observed I Zw 18 with 4 pointings. The exposure per image was 1200s. To remove the background sky spectrum in each pointing, we integrated for 600s on an ``off'' galaxy position between science exposures. We chose the nearest in time sky spectrum to scale and subtract from each science frame. The data were reduced and flux calibrated with the standard star Feige 34 using Version 1.2.1 of the KCWI Data Extraction Reduction Pipeline\footnote{https://github.com/Keck-DataReductionPipelines/KcwiDRP}
\citep{2018ApJ...864...93M}. 
The reduced cubes are astrometrically aligned to the \textit{HST} F814W image, and then placed on a common grid, with pixel size 0.15\arcsec $\times$ 0.15\arcsec, using the astronomical mosaic image engine {\tt Montage}\footnote{http://montage.ipac.caltech.edu} in combination with custom {\tt Python} scripts. A description of these steps will be presented in Rickards Vaught et al. (in prep).

\subsection{Emission Line Fitting}\label{sec:velfit}
\begin{figure*}
    \centering
    \includegraphics[width=\textwidth]{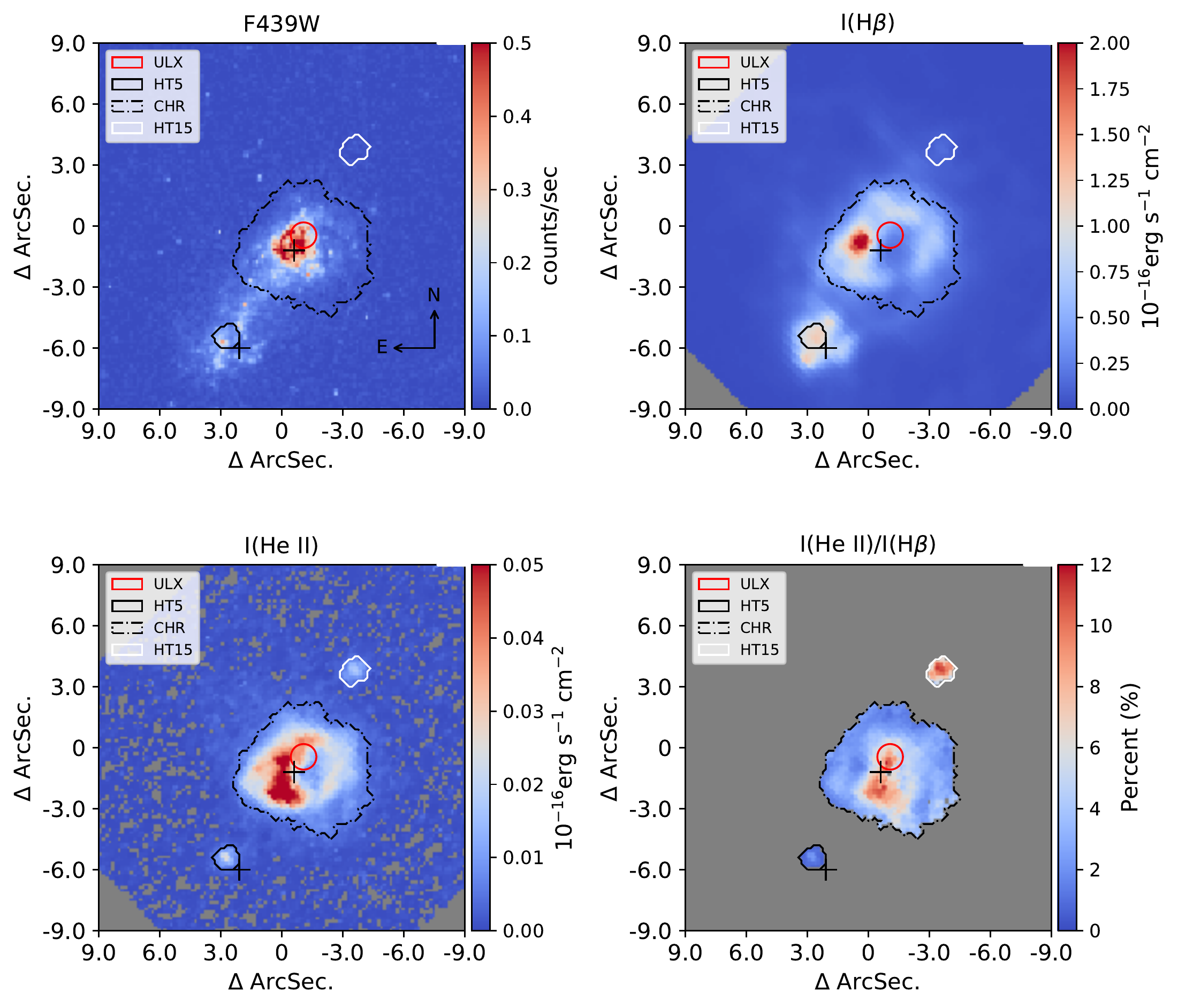}
    \caption{2D imaging of I Zw 18. \textit{Top Left:} The \textit{HST} F439W image of I Zw 18 shows lhe approximate locations of the IZW18-SE/NW (cross-hairs) in comparison to the \ion{He}{3} regions. The borders of the three \ion{He}{2} emitting regions, \SE\ (black-dashed), \MB\ (black-Solid) and the \NW\ (white-solid) are defined to contain pixels with \ion{He}{2} emission S/N $>$ 3. Also shown is the position and astrometric uncertainty of the ULX (red-solid). \textit{Top right and bottom left:} The integrated H$\beta$  and  \ion{He}{2} emission line maps. \textit{Bottom right:} \HeHr\ for the three \ion{He}{2} emitting regions.\label{fig:regions}}
\end{figure*}

\begin{figure*}
    \centering
    \includegraphics[width=1\textwidth]{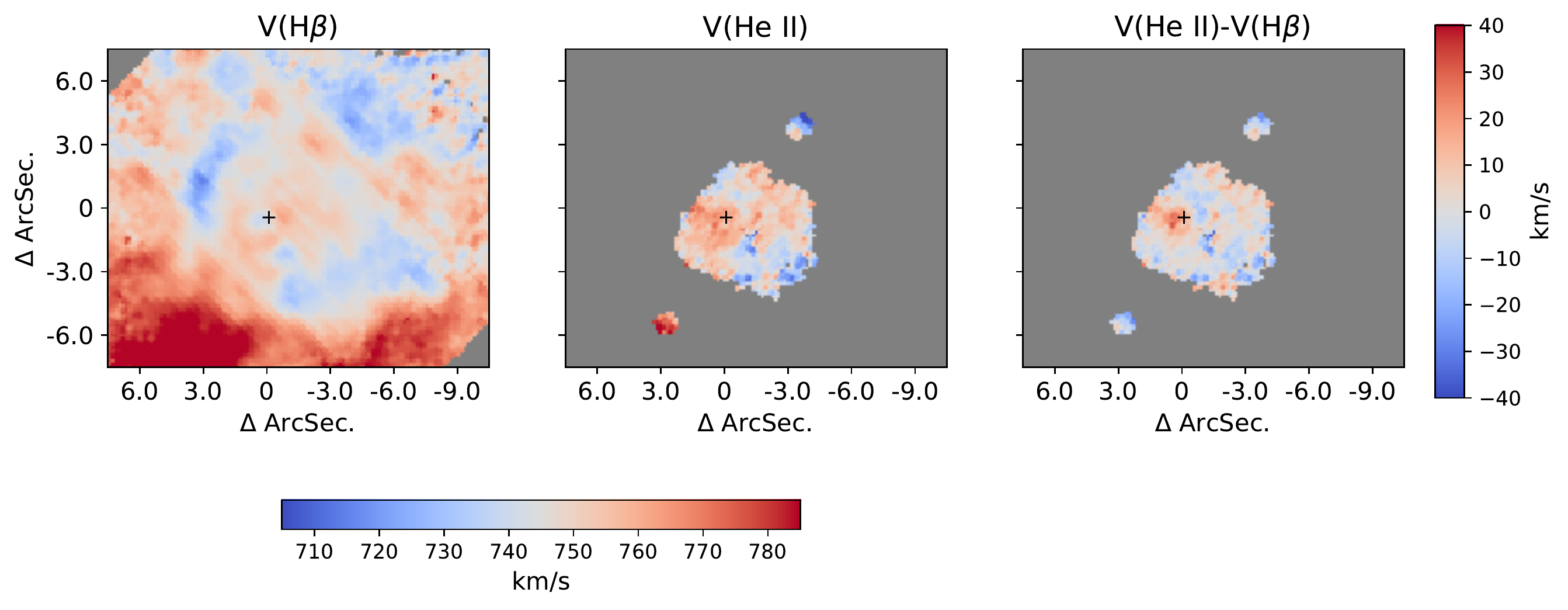}
    \caption{Two-dimensional {\tt LZIFU} velocity maps. The panels show the velocity maps measured from (left) H$\beta$, (middle) \ion{He}{2} and (right) the difference between the two velocities. The \ion{He}{2} velocities of the \SE\ and \NW\ region are very close to the H$\beta$ velocity, traced by H$\beta$, providing no clear evidence for distinct kinematics in the \ion{He}{2} emitting gas. A velocity difference of $\sim$ 30 km s$^{-1}$ can be seen near the the pixel of peak H$\beta$ emission (+), which has been attributed to a possible Supernovae Remnant \citep{Ostlin2005A&A...433..797O}.}
    \label{fig:vel}
\end{figure*}

Two-dimensional emission line maps were created using {\tt LZIFU} \citep{Ho2016}. {\tt LZIFU}  simultaneously fits a single (or multi) component Gaussian model to multiple emission lines in a spectrum. The stellar contribution is fit using an implementation of the penalized pixel fitting  routine \citep[{\tt pPFX},][]{Cappellari2004}. To determine if the observed \ion{He}{2} emission line has distinct kinematics, we perform separate {\tt LZIFU} fits specifically for \ion{He}{2} and H$\beta$ alone. These resulting maps are shown in Figures \ref{fig:regions} and \ref{fig:vel}.

\section{Astrometry of the ULX Source}
\label{sec:astrometry}
To establish how the ULX in I Zw 18 impacts the strength of \ion{He}{2} emission it is necessary to determine the location of the ULX. To achieve this, we follow the X-ray source detection procedure in \citet{Thuan2004ApJ...606..213T}\footnote{We run the CIAO, v4.12, wavelet algorithm WAVDETECT on a 0.5-10 keV image, with a probability threshold set to $10^{-7}$.}.
Next, we overlay the \textit{Chandra} source region file on top of \textit{r}-band SDSS images. We chose to match to SDSS imaging rather than \textit{HST} images because the SDSS imaging covers a larger portion of the \textit{Chandra} FoV. We are to able determine the positional offsets between the optical/X-ray positions using 5 X-Ray sources matched to SDSS optical counterparts with RMS scatter of $\sim 0.5\arcsec$. Because the astrometry of our KCWI data is anchored to the F814W \textit{HST} image, we then compute off-sets using a multitude of bright sources in both our SDSS and F814W image. The resulting RMS scatter of the offsets is $\sim0.2\arcsec$. The position of the ULX in the KCWI image is found after applying the above offsets directly to the \textit{Chandra} position of the ULX. The final co-ordinates, (9h34m1.97s, +55d34m28.33s), place the ULX near IZW18-NW, consistent with the position reported in \citet{Thuan2004ApJ...606..213T}, with RMS scatter $ 0.55\arcsec$. The position of the ULX source as it compares to H$\beta$ and \ion{He}{2} emission\footnote{The position of the ULX shown in \cite{Kehrig_2021} agree, within the positional uncertainties, with this work.} is shown in Figure \ref{fig:regions}.

\section{Results}\label{sec:results}
\subsection{3 He III Regions}
Our KCWI observations detect three \ion{He}{3} regions. First, northwest of IZW18-NW, and coincident with the \ion{H}{2} region HT15 \citep{Hunter1995ApJ...452..238H} is the \ion{He}{3} region which we designate as HT15. Next, coincident with IZW18-SE is a second \ion{He}{3} region. This region, whose \ion{He}{2} emission has been reported in \cite{Skillman1993ApJ...411..655S} and \cite{1997ApJ...487L..37I}, appears to sit close to (or on) the identified \ion{H}{2} region HT5 \citep{Hunter1995ApJ...452..238H}; as such we designate this \ion{He}{3} region as HT5. 
The last of the three we define as the Central \ion{He}{3} Region (CHR) as it is coincident with the ionized gas around IZW18-NW and in-between HT5 and HT15. CHR corresponds to the region previously mapped in \cite{Kehrig2015ApJ...801L..28K}. Our observed \ion{He}{2} emission map is shown in Figure \ref{fig:regions} with the \ion{He}{3} regions highlighted.

At a distance of 18.2 Mpc and assuming negligible reddening by dust (for more details on the very low dust attenuation in I Zw 18, see Rickards Vaught et al. in prep; \citeauthor{Cannon2002ApJ...565..931C}  \citeyear{Cannon2002ApJ...565..931C}; \citeauthor{Fisher2014Natur.505..186F} \citeyear{Fisher2014Natur.505..186F}) the total $L_{\lambda 4686}$ measured within the contours, shown in Figure \ref{fig:regions}, in the \MB\ is $(102 \pm 15.0) \times 10^{36}$ ergs s$^{-1}$. $L_{\lambda 4686}$ for the regions \SE\ and \NW\ are ($1.96\ \pm\ 0.29)\times 10^{36}$ and $(2.05\ \pm\ 0.31)\times 10^{36}$ ergs s$^{-1}$. We also report here the luminosity in H$\beta$, $L_{\text{H}\beta}$. For \MB\, $L_{\text{H}\beta}$ is $(208 \pm 43) \times 10^{37}$ ergs s$^{-1}$. $L_{\text{H}\beta}$ for \SE\ and \NW\ are $(17.7 \pm 2.6)\times 10^{37}$ and $(2.47 \pm 0.37)\times 10^{37}$ ergs s$^{-1}$ respectively. The uncertainties in the reported luminosities are dominated by an estimated calibration error of 15\%.
\SE\ and \NW\ are separated by a distance of $\sim 900$ pc and are co-linear with an axis that runs through the position of the ULX. Because of our high resolution, 0.7\arcsec\ or $\sim 60$ pc at 18.2 Mpc, we are able to resolve the morphology of the \ion{He}{2} emitting gas in the \MB\ of I Zw 18. The \ion{He}{2} and H$\beta$ emission trace a horseshoe-like shell with major/minor diameters of $\sim 550$ and $450$ pc, respectively. Nearly coincident with the ULX is an apparent cavity, likely created by stellar feedback \citep{Stasinka1999A&A...351...72S,Pequignot2008A&A...478..371P}, with projected radius, $R_C \sim 80$ pc. The \ion{He}{2} emission is preferentially extended $\sim$ 250 pc away from the ULX towards the SE.

To date, there has only been one published IFS analysis of the \ion{He}{2} emission in I Zw 18. \cite{Kehrig2015ApJ...801L..28K} spatially resolved the \ion{He}{2} emission of the \MB\ and measured a total \ion{He}{2} luminosity of ($112 \pm 7) \times 10^{36} \text{ erg s}^{-1}$; a value that is within the uncertainty reported in this work.
\cite{Kehrig2015ApJ...801L..28K} do not report any \ion{He}{2} emission near the location of \SE\ and \NW. The absence of these regions in their data is expected given their sensitivity.

\subsection{The He II($\lambda4686$)/H$\beta$ Ratio}\label{sec:He2H}
\begin{figure*}
    \centering
    \includegraphics[width=\textwidth]{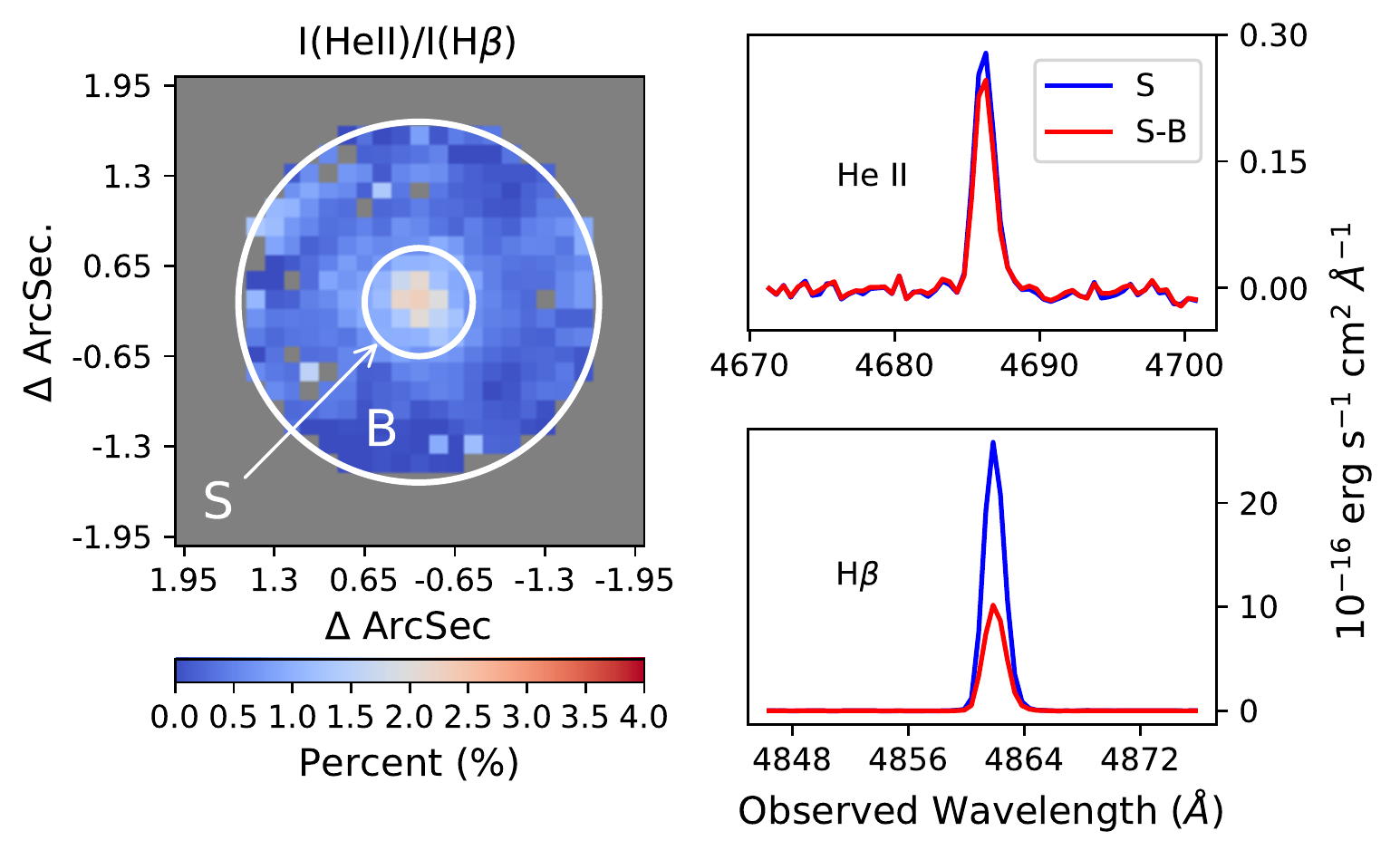}
    \caption{\HeHr\ in \SE. \textit{Left:} A $4\arcsec\times4\arcsec$ stamp of the 2D map of \HeHr\ in \SE.
    Contained in the inner annulus is the ``source'' S and contained in by the outer annulus but outside the inner annulus is the ``background'' B. 
    \textit{Right:} The two panels show the continuum subtracted integrated line profile of \ion{He}{2} and H$\beta$ of the source before (blue) and after (red) background subtraction.
    The ``background'' contributes a larger fraction to the integrated line intensity of H$\beta$ than to the \ion{He}{2} profile.}
    \label{fig:SERegionAnnulus}
\end{figure*}

The \HeHr\ intensity ratio is sensitive to shape of the Lyman continuum spectrum shortwards of 228 \AA, ionization parameter \citep{Garnett1991ApJ...373..458G,Guseva2000ApJ...531..776G,Schaerer2019A&A...622L..10S,Barrow2020MNRAS.491.4509B, Stasinka1999A&A...351...72S} and/or the shock velocity \citep{Allen2008ApJS..178...20A}. We create a \HeHr\ map 
by dividing maps of \ion{He}{2} and H$\beta$ in regions where S/N $>3$ in both lines. This map, shown in Figure \ref{fig:regions}, reveals \HeHr\ as high as 12\% in the \MB\ and \NW,  indicative of a high ionization parameter/harder ionizing spectrum object and/or shocks near the ULX and in \NW. The \ion{He}{2} emission along the eastern edge of the shell, and the emission co-spatial with the ULX, exhibits the largest \HeHr\ enhancement. 
Compared to \NW\ and the \MB, \SE\ exhibits a low peak \HeHr\ value of $\sim 2\%$. Because \SE\ is located within IZW18-SE, there may be excess H$\beta$ emission from gas ionized by stellar sources contributing to the spectrum of \SE. To remove the effects of such an ionized gas component we subtract the median local spectrum in an annulus between 0.5-1\arcsec\ surrounding \SE. The results of this subtraction are shown in Figure \ref{fig:SERegionAnnulus}; the spectrum of \SE\ with the local spectrum removed is revealed to have \HeHr\ closer to $\sim$ 4.5\%.

\subsection{Undetected Stellar Continuum in \NW}\label{sec:nwlimit}
\begin{figure*}
    \centering
    \includegraphics[width=\textwidth]{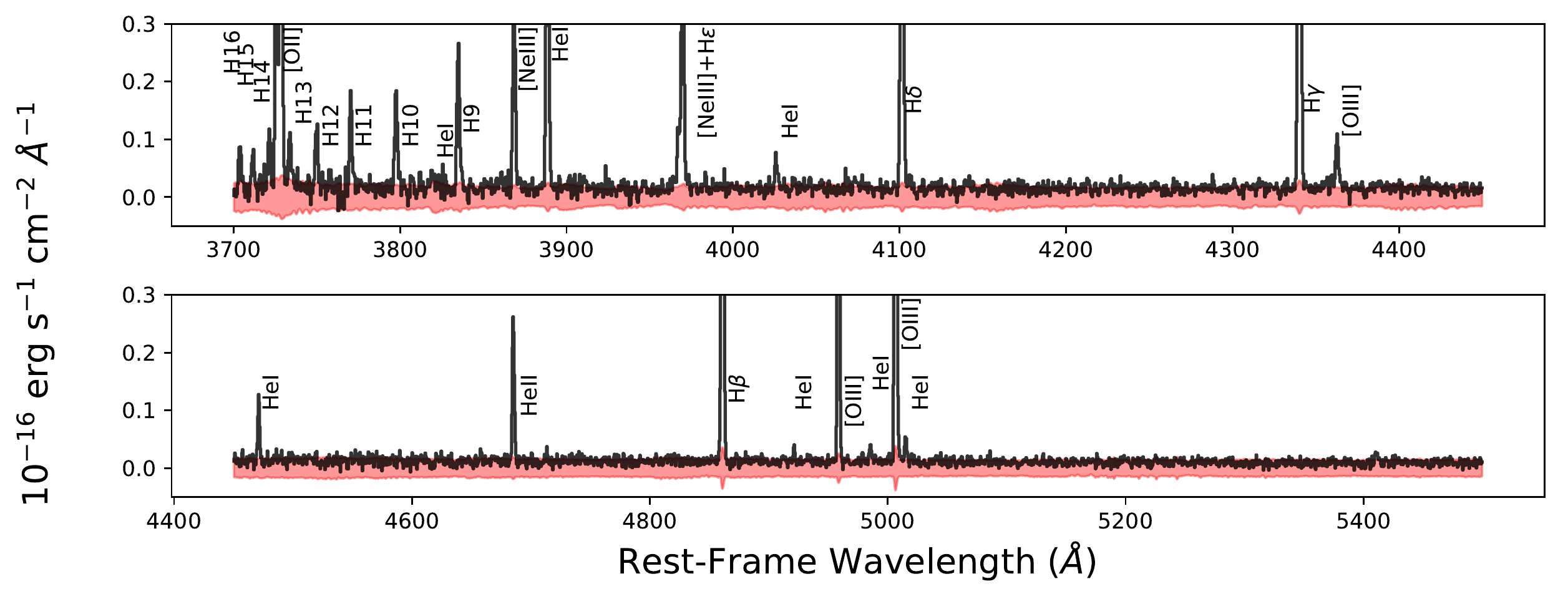}
    \includegraphics[width=1\textwidth]{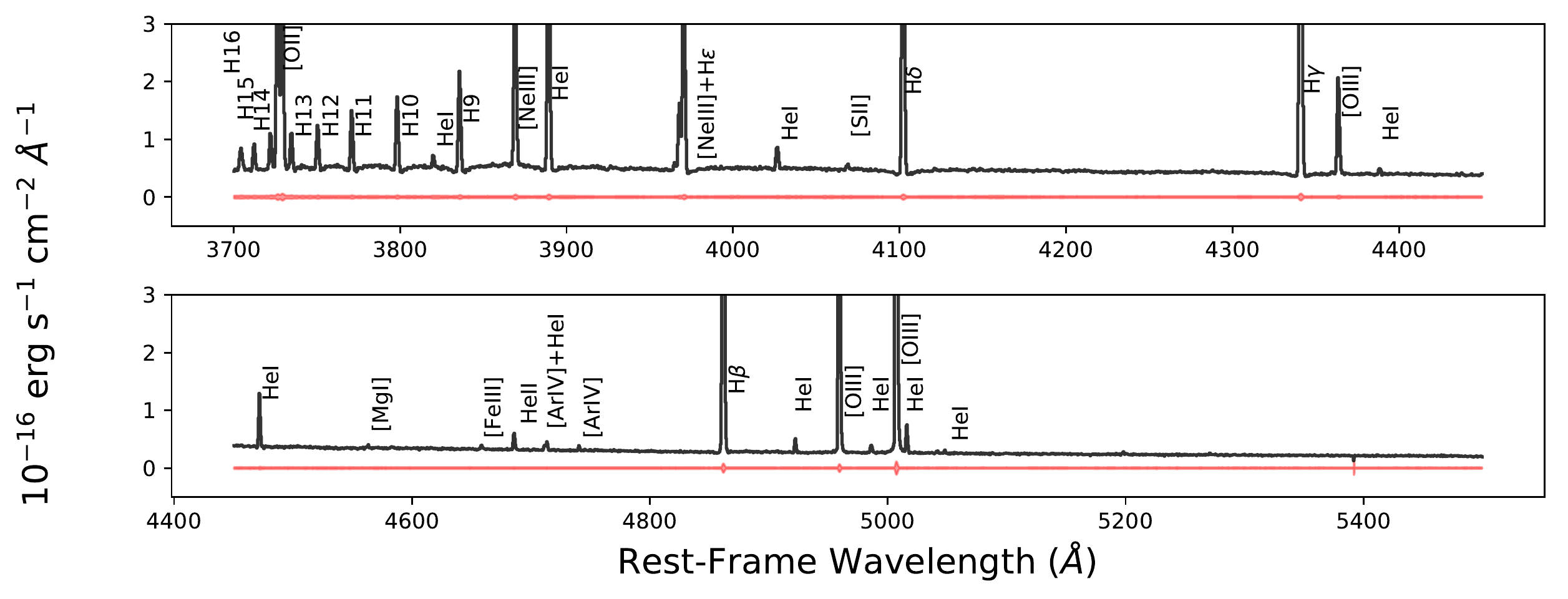}
    \caption{The integrated spectrum of \NW\ and \SE\ (labels of lines shortward/longward of [\ion{O}{2}] are shifted 3 \AA\ to the left/right for clarity). \textit{Top:} The integrated spectrum of \NW\ (black) is shown with the error (red). The continuum flux density of \NW\ region is observed to be within the instrumental noise along the full wavelength range. We infer from this that the \NW\ region stellar population is not detected in these data.
    \textit{Bottom:} Integrated spectrum of \SE. The highest ionization line in either of these spectra is \ion{He}{2}($\lambda$4686).}
    \label{fig:NWspectrum}
\end{figure*}

Of the three regions, \NW\ has a non-detected stellar continuum in our observations. This can be seen in Figure \ref{fig:NWspectrum}, where we plot the integrated spectrum of \NW\ measured using a $\sim 1$ arcsec$^2$ aperture covering \NW. After adding in quadrature the $\pm$1-$\sigma$ error spectrum for each pixel, we find that the continuum flux across all wavelengths is within the integrated 1-$\sigma$ errors. If a stellar contribution to the continuum were present, it is undetected below a representative $\bar{\sigma}$ $\approx1.3\times10^{-18}$ \fluxu, where $\bar{\sigma}$ is the median 1-$\sigma$ error across all wavelengths. This flux corresponds to a limiting apparent magnitude of $m_{V} > 24$. This result is similar to those of \citet{Hunter1995ApJ...452..238H} and \citet{Hunt2003ApJ...588..281H}, whose H$\alpha$ and NIR observations of \NW\ also lack a measurable stellar component, but instead, only show emission lines from ionized gas. Furthermore, the Hubble Source Catalog \citep{Whitmore2016AJ....151..134W} classification of object, 766559, at the position of \NW\ suggests an extended object, rather than a point source according to photometry in the filters: F450W, F555W, F702W, F814W.

\subsection{Kinematics of the He II gas}\label{sec:kinematics}
To test whether or not the kinematics of \ion{He}{2} gas are distinct from that of H$\beta$, as might be expected if \ion{He}{2} emission is the result of shocks while H$\beta$ is largely from photoionization, we perform single line {\tt LZIFU} fits. The results/comparisons of the velocity fits are shown in Figure \ref{fig:vel}. We find that the difference between the \ion{He}{2} and H$\beta$ velocities in the \NW/ \SE\ regions is negligible, signifying that dynamics of the \ion{He}{2}-emitting gas around \NW\ and \SE\ have velocities similar to those traced by H$\beta$. We measure a velocity difference of $\sim 30$ km s$^{-1}$ near the position of peak H$\beta$ emission, however this may be associated with a supernova remnant or young stellar cluster \citep{Ostlin2005A&A...433..797O}.

\section{Discussion}\label{sec:discussion} 
We use our measurements of the spatial distribution of \ion{He}{2} emission in I Zw 18 to discuss possible origins of the high ionization state gas in the two newly resolved \ion{He}{3} regions. This includes Wolf-Rayet stars and ULX generated phenomena. In the ULX case, the alignment of the two \ion{He}{3} regions with the ULX source leads us to explore the possibility that \NW\ and \SE\ are observable effects from jet/outflow generated shocks or beamed X-ray emission originating from the ULX. But first, we discuss if Wolf-Rayet stars are consistent with the properties of the \ion{He}{2} emission in \NW\ and \SE. 

\subsection{Wolf-Rayet Stars}
\label{sec:stellar_source}
Although the spectra of \SE\ and \NW\ lack broad-lined, stellar \ion{He}{2} emission, a signature of WR stars \citep{Crowther2006A&A...449..711C}, we are unable to eliminate these objects as the source of \ion{He}{2} emission.
The absence of broad-lined, stellar \ion{He}{2} emission in our data would be consistent with a population of nitrogen rich WR (WN) stars, as modeling of WN atmospheres in \citet{Crowther2006A&A...449..711C} has demonstrated that WN winds are reduced at sub Large and Small Magellanic Clouds (LMC/SMC) metallicities. Reduced winds diminish the production of broad-lined \ion{He}{2} emission, and could explain the absence of this feature in these regions while still potentially powering nebular \ion{He}{2} emission.

\ion{He}{3} regions have been observed around WN stars Brey2 and AB7 in the LMC and SMC respectively \citep{Naze2003A&A...401L..13N,Naze2003A&A...408..171N}. In \citet{Naze2003A&A...401L..13N} the \ion{He}{3} region around Brey2 has a \ion{He}{2} luminosity, $L_{\lambda 4686}= 3.5\times10^{35}$ erg s$^{-1}$. For the AB7 \ion{He}{3} region, $L_{\lambda 4686}$ is 10$\times$ greater than Brey2 \citep{Naze2003A&A...408..171N}. Both objects exhibit regions with \HeHr\ greater than 10\%. Compared to our \NW\ and \SE\ $L_{\lambda 4686}$ budget, $6$ Brey2-like stars could account for the \ion{He}{2} luminosities in \NW\ and \SE\ while, to order of magnitude, a single AB7-like star could solely power \ion{He}{2} emission in both \NW\ and \SE. Applying the distance modulus to the absolute magnitude, $M_V\sim -3$, for WN stars \citep{Crowther2006A&A...449..711C} these populations wouldn't appear brighter than $m_V \sim 24$. Compared to these He$^+$ ionizing WN2-4 sub-types, brighter WN5-10 stars are too cool and are unable to doubly ionize Helium. WN stars are viable candidates for the production of \ion{He}{2} emission in \NW\ and \SE. 

\subsection{Jet or Beamed X-ray powered He II emission in I Zw 18}
\label{sec:JetPower}
Supposing that the alignment of \NW\ and \SE\ with the ULX arises from a physical link, then jet/outflow generated shocks or beamed X-rays originating from the ULX would be needed to explain the alignment and emission. 

Various collisional and radiative processes generated behind shock fronts have been modeled by \citet{Allen2008ApJS..178...20A}. This modeling shows that shock speeds between $100-150$ km s$^{-1}$
can produce \HeHr\ ratios consistent with those observed for \NW\ and \SE. Because shock-sensitive lines such as [\ion{O}{1}]($\lambda6300$), [\ion{N}{2}]($\lambda6584$), [\ion{S}{2}]($\lambda6717,6731$) and [\ion{Ne}{5}]($\lambda3346,3426$) lie outside our spectral range, we are unable to compare to those with shock velocity estimates inferred from \HeHr. \citet{Kehrig2016MNRAS.459.2992K} report measurements of these lines, excluding [\ion{Ne}{5}], for the \MB. The observed line ratios are found to be inconsistent with shock ionization. We note that shock templates at I Zw 18-like metallicities are uncertain as shock models only exist for metallicities $\geq$ SMC.
Moreover, the difference between the \ion{He}{2} and H$\beta$ derived velocities, as shown in Figure \ref{fig:vel}, does not show evidence of $\sim 100$ km s$^{-1}$ jet/outflow generated shocks at the locations of \NW\ and \SE. The absence of such velocities does not necessarily discount the possibility of jet/outflow activity if the motion were primarily in the plane of the sky. Low spatial resolution radio observations of I Zw 18 by \cite{Hunt2005A&A...436..837H} show an extended synchotron halo + lobe structure in the radio continuum which they take as evidence of a wind-blown super-bubble accompanied by bi-polar outflows. However, the direction of these outflows is perpendicular to the axis connecting \NW\ and \SE. 

Next, we test whether isotropic X-ray emission from the ULX could be sufficient to power the He II emission, or if beaming would be needed. The X-ray flux, using $L_{X}=1\times10^{40}$ erg s$^{-1}$, at a distance of, $R \sim 450$ pc, is $F_X\sim4\times10^{33}$ erg s$^{-1}$ pc$^2$. The power passing through a surface area of $\pi r^2$, where $r$ is the radial size of \NW/\SE\ with the value $r=0.3$\arcsec\ or 30 pc, is $P_X=F_X \times \pi r^2\sim1\times10^{37}$ erg s$^{-1}$. 

As observed in a number of cases and reproduced by photoionization modeling calculations, X-ray ionized nebulae around ULXs appear to exhibit \ion{He}{2} emission to total X-ray luminosity ratios of $L_{\lambda 4686}/L_X \sim 10^{-4}$ \citep{Pakull1986Natur.322..511P,Pakull2002astro.ph..2488P, Kaaret2004MNRAS.351L..83K,kaaret2009ApJ...697..950K,Moon2011ApJ...731L..32M}.
Assuming this same fraction of power goes into producing \ion{He}{2} emission in HT5 and HT15, the power for \ion{He}{2} emission is $P_{\lambda 4686}=P_X\times 10^{-4}\sim 10^{33}$ ergs s$^{-1}$. When compared to the observed $L_{\lambda 4686}$ for \NW/\SE\ the \ion{He}{2} production budget is short by orders of magnitude. Even using $L_{\lambda 4686}/L_X \sim 10^{-2}$, which one gets assuming the ULX is responsible for all of the \ion{He}{2} emission in the \MB, the power available to produce \ion{He}{2} emission is short by two orders of magnitude. This shows that if the X-ray emission from the ULX powers these two \ion{He}{3} regions, the emission needs to be beamed rather than isotropic.

\section{Conclusion}\label{sec:conclusions}
We presented KCWI observations of I Zw 18. Our observations revealed the presence of two \ion{He}{3} regions, \NW\ and \SE, in addition to the \ion{He}{3} region mapped by \citet{Kehrig2015ApJ...801L..28K}. Enhanced \HeHr\ ratios between 4\% and up to 12\% are measured in \SE\ and \NW. Region \NW, which shows some of highest \HeHr\ values, has an undetected stellar population ($m_V > 24$). We compared the observed \ion{He}{2} luminosity in \NW\ and \SE\ to \ion{He}{3} regions surrounding LMC/SMC WN stars and find that similar objects are sufficient to produce the \ion{He}{2} luminosity and \HeHr\ enhancement of \NW\ and \SE\ as well as the absence of broad-lined, spectral \ion{He}{2} features, whilst remaining below our detection limit.

Based on the alignment of the two \ion{He}{3} regions and the ULX, we explored a scenario in which jet/outflow activity or beamed X-ray emission originating from the ULX powers the observed \ion{He}{2} emission in \NW\ and \SE. Due to our spectral coverage and the lack of shock models appropriate for the galaxy's metallicity, we cannot put a strong constraint on whether shocks could be powering \ion{He}{2} emission. We assessed the velocity structure of the ionized gas and found no kinematic anomalies driven by jet/outflow activity. Assuming that \NW\ and \SE\ are illuminated by isotropic X-ray emission, we found that the ULX would not produce sufficient X-ray flux to generate the observed \ion{He}{2} emission. If the X-ray emission from the ULX powers these sources, it would require beaming.
We will present further analysis of the metallicity and temperature structure of I Zw 18 in an upcoming publication.

\acknowledgements
The authors thank the referee for very thorough and enlightening reports which significantly improved the analysis presented in this letter. The authors also thank Brent Groves for useful conversations. The data presented herein were obtained at the W. M. Keck Observatory, which is operated as a scientific partnership among the California Institute of Technology, the University of California and the National Aeronautics and Space Administration. The Observatory was made possible by the generous financial support of the W. M. Keck Foundation. The authors wish to recognize and acknowledge the very significant cultural role and reverence that the summit of Maunakea has always had within the indigenous Hawaiian community.  We are most fortunate to have the opportunity to conduct observations from this mountain. We also wish to thank Luca Rizzi and all the Keck Observatory staff for observational support.
RRV and KS acknowledge funding support from National Science Foundation Award No.\ 1816462.
This research made use of Montage. It is funded by the National Science Foundation under Grant Number ACI-1440620, and was previously funded by the National Aeronautics and Space Administration's Earth Science Technology Office, Computation Technologies Project, under Cooperative Agreement Number NCC5-626 between NASA and the California Institute of Technology.
Funding for the Sloan Digital Sky Survey IV has been provided by the Alfred P. Sloan Foundation, the U.S. Department of Energy Office of Science, and the Participating Institutions.

\bibliography{RRickardsVaughtHeII}
\bibliographystyle{aasjournal}
\end{document}